\documentstyle[aps,prl,twocolumn,epsf]{revtex}

\newcommand\bxi{\mbox{\boldmath $\xi$}}
\newcommand\bsig{\mbox{\boldmath $\sigma$}}
\newcommand\bcv{\mbox{\boldmath ${\cal V}$}}

\newcommand \ltdash{\raise-1.8pt\hbox{$\scriptscriptstyle |$}}

\newcommand \bea {\begin{eqnarray} }
\newcommand \eea {\end{eqnarray}}

\newcommand \dg {^{\dagger}}

\newlength{\bxwidth}\bxwidth=0.8\textwidth

\begin{document}
\title{Oscillatory instabilities in d.c.\ biased quantum dots}

\author{P. Coleman$^1$, C. Hooley$^{1,2}$, Y. Avishai$^3$, Y. Goldin$^3$, and A. F. Ho$^4$}

\address{$^1$Center for Materials Theory,
Rutgers University, Piscataway, NJ 08854, U.S.A.\\
$^2$School of Physics and Astronomy, Birmingham University, Edgbaston, Birmingham B15 2TT, U.K.\\
$^3$Department of Physics, Ben Gurion University, Beer
Sheva, Israel\\
$^4$Department of Physics, 
Oxford University, 
1 Keble Road, 
Oxford OX1 3NP, U.K.}

\address{\rm (Received: )}
\address{\mbox{ }}
\address{\parbox{14cm}{\rm \mbox{ }\mbox{ }
We consider a `quantum dot'
in the Coulomb blockade regime, subject to an arbitrarily large
source-drain voltage $V$.  When $V$ is small, quantum dots
with odd electron occupation display the Kondo
effect, giving rise to enhanced conductance.  Here we investigate the
regime where $V$ is increased beyond the Kondo temperature and 
the Kondo resonance splits into two components.  It is shown that interference
between them results in spontaneous oscillations of the current
through the dot.  The theory predicts the appearance of ``Shapiro steps''
in the current-voltage characteristics of an irradiated quantum dot; these
would constitute an experimental signature of the predicted effect.
}}
\address{\mbox{ }}
\address{\parbox{14cm}{\rm PACS No: 73.63.Kv, 72.10.Fk}}
\maketitle

\makeatletter
\global\@specialpagefalse
\makeatother

\narrowtext

\par
Quantum dot
devices \cite{Kastner:1992} display
very rich physics in and near equilibrium.  
In particular, experiments
\cite{Goldhaber:1998}
over the past few years 
have confirmed a decade-old theoretical prediction
that these systems exhibit 
the Kondo effect \cite{Ng:1988}:\ the low temperature 
formation of a hybridization resonance 
between
the dot and the leads to which it is connected.  This resonance
is the result of collective spin exchange processes between the leads and the dot,
which dominate the low temperature physics below a certain scale (the
`Kondo temperature').  The Kondo effect is manifested as an enhancement of
the dot's conductance.
\par
A bias voltage applied to a quantum dot modifies the electron energies
in the leads, driving a current through the dot. This offers a unique
opportunity to study correlated electrons out of equilibrium.
Can new kinds of non-equilibrium collective phenomena develop in
driven quantum systems?  In the realm of classical physics, there are
many instances of transitions to new phases in response to a
non-equilibrium driving force. An example is the Rayleigh-B{\'e}nard
instability, where a temperature difference causes the development
of convective roll patterns that spontaneously select their own
spatial and temporal frequencies \cite{Meyer:1999}. However, for a driven
quantum system to develop new collective behavior, it must preserve
its quantum mechanical coherence. The issue of
whether a bias voltage preserves the phase coherence of a quantum dot
is thus a matter of considerable concern.
\par
\begin{figure}
\begin{center}
\leavevmode
\hbox{\epsfxsize=8cm \epsffile{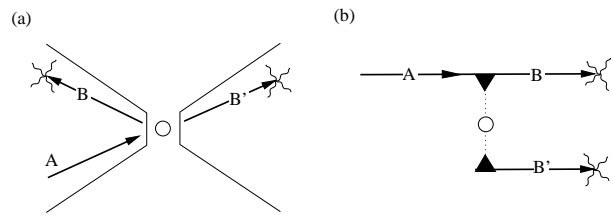}}
\end{center}
\caption{(a) A schematic depiction of an
incoming electron wave A scattering off the quantum dot into the same (B) and
the opposite (B') leads.  Dephasing (indicated by wavy crosses) takes place after
the scattering event.  (b) The mapping of the quantum
dot problem to a one dimensional ``chiral'' model.
Electrons and information  travel from left to right.  Dephasing of
electrons at B and B' cannot affect the quantum dot.}
\label{fig1}
\end{figure}

In this paper we argue that electron flow through a quantum dot
modifies, but does not dephase the correlations of 
the dot.  Central to our
arguments is the observation that the physics of a quantum
dot can be mapped onto a one-dimensional problem \cite{Chiral}, where the
electrons in each 
lead are represented by waves traveling in one direction 
along one-dimensional
conductors, arriving at the dot 
from the left and leaving it to the right (Fig.~\ref{fig1}). 
This chiral mapping is exact. 
Electrons scattering off the
dot do not lose phase information until they suffer subsequent
inelastic collisions with phonons or magnetic impurities in the leads, 
which, in the chiral mapping picture, lie to the `right' of the dot. Since
information travels only from left to right, the dephasing of an
outgoing electron does not affect the spin it has left behind, and
thus a current of electrons through a quantum dot does not dephase the
Kondo effect, but acts as a coherent driving force.
\par
This argument is employed to examine the possibility
of dynamical instabilities in a driven, yet fully
coherent quantum dot.
We present calculations which predict that, beyond a
critical bias voltage $V_{c}$, the current flowing through the quantum
dot spontaneously acquires an oscillatory component whose frequency is
a non-linear function of the bias voltage. 
\par
\begin{figure}
\begin{center}
\leavevmode
\hbox{\epsfxsize=6.5cm \epsffile{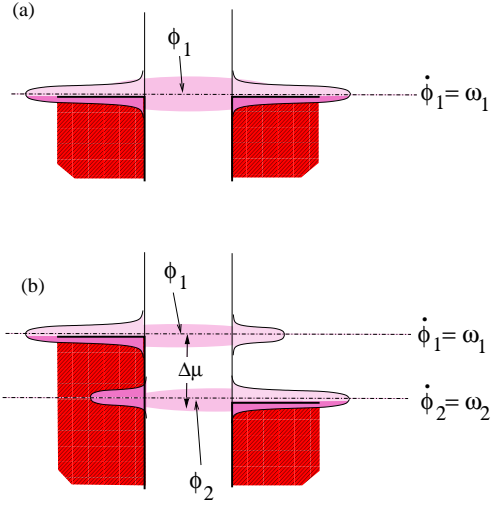}}
\end{center}
\caption{(a) Schematic picture of the equilibrium situation:\ a single
Kondo resonance is formed between the electron
on the dot and electrons in both leads.
(b) Quantum dot under a DC bias $V > V_c$:\ Kondo resonances develop
at each chemical potential, but {\em each\/} resonance still involves the
electrons from {\em both\/} leads.  Interference between the 
resonances drives the oscillatory current.}
\label{fig2}
\end{figure}

%
The quantum dot system considered here consists essentially of three parts:\
the dot itself, and the left and right leads, to which it is connected
by weak tunneling junctions.  In the Coulomb
blockade regime, the occupation of the dot is fixed, and the only allowed processes involve
flipping the spin of the unpaired electron on the dot.  In this simple
picture, a dot containing an odd number
of electrons is a spin-${1 \over 2}$ impurity.
As the temperature is lowered, spin fluctuations lead to the
development of a sharp resonance peak in the interacting electron
density of states:\ the Abrikosov-Suhl (AS) resonance.  This peak occurs
at an energy $\hbar \omega_1$, close to the common Fermi energy of
the leads, and has a width $\Delta \sim T_K$, the Kondo temperature.
(See Fig.~\ref{fig2}(a).) 
\par
In the description which we shall use, 
the coherence associated with the Kondo effect is encoded in terms of
a time-dependent hybridization field ${\hat \Gamma}_\beta(t) \sim
\sum\limits_{{\bf k}\sigma} c^{\dg}_{\beta{\bf k}\sigma} d_\sigma$ between
the lead electrons and the electron on the dot.  (The operator
$c^{\dg}_{\beta{\bf k}\sigma}$ creates a lead electron with momentum ${\bf k}$
and spin $\sigma$ in lead $\beta \in \left\lbrace L,R \right\rbrace$;
$d_\sigma$ annihilates a dot electron with spin
$\sigma$.)  Within the mean field formalism (see below), ${\hat\Gamma}_\beta$
is replaced by its mean value $\Gamma \equiv \langle {\hat\Gamma}_\beta \rangle
= \Gamma_0 e^{i\omega_1 t}$.
\par
In the presence of a sufficiently large bias voltage $V$, the AS
resonance splits into two
peaks \cite{Hershfield:1991}, one
near each Fermi level.  (See
Fig.~\ref{fig2}(b).)  $\Gamma(t)$ therefore becomes
a sum of two terms, $\Gamma(t) = \Gamma_1 e^{i\omega_1 t}
+ \Gamma_2 e^{i\omega_2 t}$, where $\omega_1$ and $\omega_2$ are
the level positions of the two AS peaks.  Quantum interference
between these two terms produces an oscillatory current at
frequency $(\omega_1-\omega_2) \sim eV/\hbar$.
We emphasise that this oscillating current originates as
a response to a DC bias:\ this is to be contrasted with the case
where the driving voltage is itself
time-dependent \cite{Jauho:1994}.
\par
%
A natural choice for the description of a quantum dot coupled to two
leads is the Anderson model.  We consider a generalized version of such a model,
allowing for an asymmetry between transmission and reflection
amplitudes (see below).  In the Coulomb blockade regime, charge fluctuations
are virtual, and may be integrated out \cite{Schrieffer:1966} to
obtain an effective spin-exchange Hamiltonian:
\bea
H & = & H_0 + H_R + H_T; \label{ham} \\
H_0 & = & \sum_{\beta{\bf k}\sigma} \left( \epsilon_{\bf k} - \mu_\beta \right)
c^{\dg}_{\beta{\bf k}\sigma} c_{\beta{\bf k}\sigma}, \nonumber \\
H_R & = & - {J \over 2N} \sum_{\sigma\tau} \left[
c^{\dg}_{L\sigma} d_\sigma d^{\dg}_\tau c_{L\tau} + (L \to R) \right], \nonumber \\
H_T & = & - (1-\eta) {J \over 2N} \sum_{\sigma\tau} \left[
c^{\dg}_{R\sigma} d_\sigma d^{\dg}_\tau c_{L\tau} + (L \leftrightarrow R) \right], \nonumber
\eea
where $c^{\dg}_{\beta\sigma} \equiv \sum\limits_{\bf k} c^{\dg}_{\beta {\bf k}\sigma}$.
The spin variables $\sigma,\tau$ now take integer values from $1$ to $N$, a generalization
which we discuss below.
Here $H_0$ controls the electrons in the leads, with 
$\mu_L\,{-}\,\mu_R\,{\equiv}\,eV$, the source-drain voltage.
The remaining parts of the Hamiltonian describe the spin exchange processes which accompany
electron reflection ($H_R$) and transmission ($H_T$) at the dot.
\par
A crucial ingredient is the asymmetry parameter $\eta$.  When (\ref{ham})
is derived directly from the Anderson model, one finds that $\eta=0$; this reflects a
symmetry between reflection and transmission processes that will never be
exactly realized in a physical quantum dot system \cite{Coleman:1998}.  Moreover, it has been
proposed \cite{Coleman:2000} that, at voltages above the Kondo temperature, the
dot system develops a divergent susceptibility to perturbations of the form
$-\lambda O^{\dg}O$, where
operator $O$ is defined by $O(t) \equiv \sum\limits_{\sigma\tau} \left( 
c^{\dg}_{L\sigma} \bsig_{\sigma\tau} c_{R\tau} \right) \cdot {\bf S}$,
where ${\bf S}$ is the impurity spin.  The inclusion of
a small non-zero $\eta$ represents the effect of such terms.
\par
Our analysis of the model (\ref{ham}) is based on the self-consistent determination of the
hybridization fields ${\hat\Gamma}_{\beta\sigma}(t) = c^{\dg}_{\beta\sigma}(t)
d_\sigma(t)$.
In an actual physical system, the spin index $\sigma$ may take one of two
values ($\uparrow$ or $\downarrow$), but it is useful to consider a general
case where $\sigma \in \left\lbrace 1,\ldots,N \right\rbrace$.  This is
because, as $N \to \infty$, the hybridization fields ${\hat\Gamma}_{\beta\sigma}(t)$
behave as well defined semi-classical variables:\ ${\hat\Gamma}_\beta(t)
\to {1 \over N} \sum\limits_\sigma \left\langle c^{\dg}_{\beta\sigma}(t) d_\sigma(t)
\right\rangle$. The use of the large-$N$ approach deserves 
particular discussion.  By making this choice we render
the problem exactly solvable, but in doing so, we exclude inelastic
scattering processes and thus emphasize the coherent
aspects of the quantum dot physics.  (Specifically, the lowest order processes
that produce a lifetime for the $d$-fermion are of $O(1/N)$, and therefore
disappear in the $N\to\infty$ limit.)
The phase coherent
physics of the
$N=2$ Kondo effect is  captured by the large-$N$ limit \cite{Read:1983}.
Thus,
if an applied voltage
acts as a phase coherent driving force on the quantum dot,
we expect the large-$N$ mean field theory to correctly capture any collective
behavior that develops at large voltage bias. 
Support for this approach is provided by
recent
calculations \cite{Coleman:2000}, which show that the quantum 
dot system remains a non-perturbative phenomenon at arbitrarily large
voltages.  However, the ultimate test of the procedure must surely lie
in experiment.
\par
Replacing the hybridization fields by their expectation
values in the large-$N$ limit results in the mean-field Hamiltonian
\begin{eqnarray}
H & = & H_{\rm o} + \sum_{\sigma} \left[ ({\cal V}^*_{L} (t) c\dg_{L\sigma} d_{\sigma}
+ d\dg_{\sigma} c_{L\sigma} {\cal V}_{L} (t))+ (L\to R) ) \right] \nonumber \\
& & \quad + {N \vert {\cal V}_L + {\cal V}_R \vert^2 \over J(2-\eta)}
+ {N \vert {\cal V}_L - {\cal V}_R \vert^2 \over J\eta},
\end{eqnarray}
where ${\cal V}_{L}$ and ${\cal V}_{R}$
are defined by
\begin{equation}\label{rel}
\left(\matrix{{\cal V}_{L} (t)\cr {\cal V}_{R} (t)} \right)=
\frac{1}{2}\left[\matrix{J& J(1-\eta)\cr
J(1-\eta)&J\cr} \right]\left(\matrix{\Gamma _{L} (t)\cr
\Gamma _{R} (t)} \right).
\end{equation}
The amplitudes ${\cal V}_{L,R}(t)$ must
be self-consistently determined from the expectation values
${\Gamma_{L,R}(t)}$ at finite voltage $V$.
\par
Our analysis uses the
Schwinger-Keldysh formalism \cite{Schwinger:1961},
and from the practical point of view this requires an Ansatz for the
form of the hybridization fields ${\cal V}_\beta(t)$.  
In the present work,
two Ans{\"a}tze are investigated:\ a {\em two-frequency Ansatz\/} (``2F''), 
\begin{eqnarray}
\bcv \equiv \pmatrix{{\cal V}_L \cr {\cal V}_R \cr} & = &
{\cal V}_{\rm o} \left( \bxi_{+} e^{-i \omega_{\rm o} t} 
+ \bxi_{-} e^{i \omega_{\rm o} t}
\right), \label{ansatz} \end{eqnarray}
and a {\em one-frequency Ansatz\/}
(``1F''), where the second term on the RHS of 
(\ref{ansatz}) is absent. The 1F solutions correspond
to states with a single AS resonance.  The 2F
solutions correspond to states in which the resonance has been
split into two components at energies $\pm \omega_{\rm o}$,
giving rise to an oscillatory current. 
\par
We supplement this analytic approach with
numerical simulations, using the Heisenberg equations of motion to time-evolve
the averages
$\left\langle c^\dag_{\beta \sigma}(t) d_{\sigma}(t) \right\rangle$ and
$\left\langle c^\dag_{\alpha \sigma}(t) c_{\beta \sigma}(t) \right\rangle$, which together
describe the state of the whole system.
The initial state is taken to be a `high-temperature' state, i.e.\ one with extremely
small hybridization between the dot and the leads.
The fields ${\cal V}_\beta(t)$
are updated at each time step to ensure self-consistency.
The quantity of interest is the physical current
through the dot,
$I(t) = 
2J(1-\eta) \, \mbox{Im} \left(\Gamma_{R} (t)\Gamma_{L}^{*} (t)\right)$.
%
\par
\begin{figure}
\begin{center}
\leavevmode
\hbox{\epsfxsize=\columnwidth \epsffile{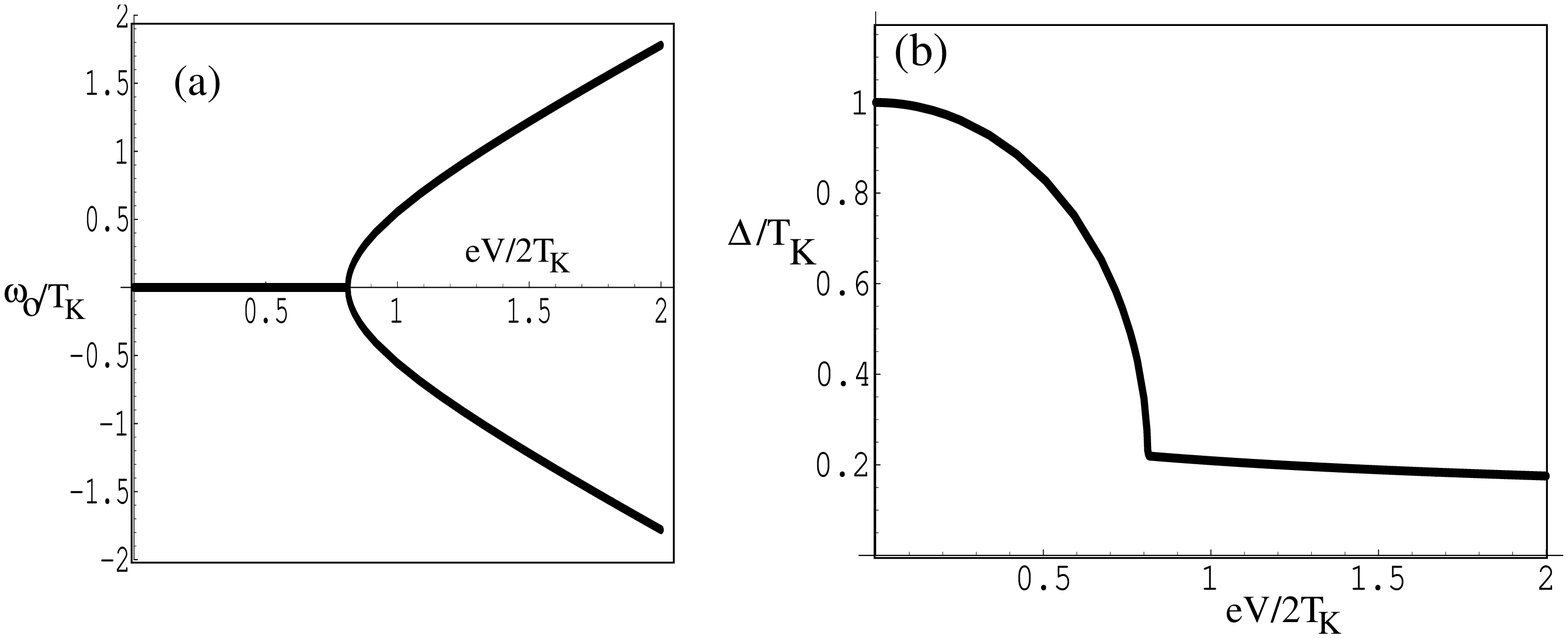}}
\end{center}
\caption{(a) The unstable frequency $\omega_{\rm o}(V)$
governing the splitting of the AS resonance, in the case
$\eta \approx 0.37$.
(b) The width $\Delta(V)$ of the AS resonance for the case
$\eta \approx 0.37$.}
\label{fig3}
\end{figure}

Consider first
the 1F solutions. These are
{\em static solutions\/} since, with only one AS resonance present, there is
no possibility of interference: the
current through the dot is hence time-independent.  
As long as $\eta \ne 0$, static solutions
exist for all values of $V$.
As $V$ is increased from zero, the level position
$\omega_{\rm o}=0$ {\em does not change\/}, i.e.\ the application of a small
bias neither splits nor shifts 
the AS peak.
The width of the peak, $\Delta$, decreases sharply as
one increases the voltage (Fig.~\ref{fig3}(b)).
At $V=V_c \approx 
2 T_K / e$, the single static solution at frequency
$\omega_{\rm o}$ bifurcates, and for $V>V_c$ one finds two {\em degenerate\/}
static solutions, with frequencies $\pm \omega_{\rm o}$, where
$\omega_{\rm o}= {eV \over 2\hbar} \, \Phi(eV/2 k_B T_K)$
(Fig.~\ref{fig3}(a)). At large voltages, $\Phi(x) \to 1$; thus
these degenerate solutions represent an AS resonance, of width
$\Delta(\eta,V)$,
attached to the left or right lead.
\par
\begin{figure}
\begin{center}
\leavevmode
\hbox{\epsfxsize=7cm \epsffile{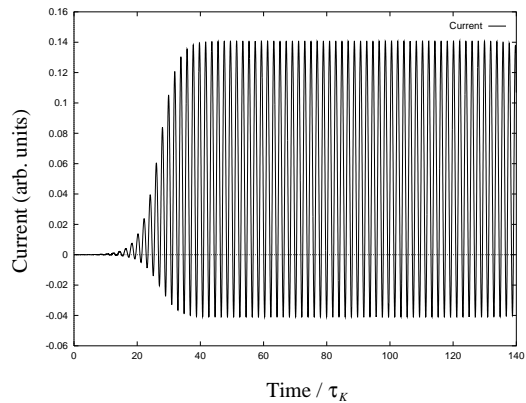}}
\end{center}
\caption{Tunneling current through the dot, $I(t)$, as a function of time, for
DC voltage $V = 3.5 \, T_K$.  (Here $\eta \approx 0.46$, and time
is measured in units of the Kondo time, ${\tau}_K \equiv \hbar / k_B T_K$.)}
\label{fig4}
\end{figure}

When $V<V_c$, the numerical simulations support 
these results; however, for
$V>V_c$, they indicate that the static solutions 
are unstable
to the emergence of a 2F solution (\ref{ansatz}), consisting
of components at both frequencies $\pm \omega_{\rm o}$.  
Quantum interference between these
two AS resonances contributes an 
oscillating part to the current $I(t)$ (Fig.~\ref{fig4}).
The oscillations are monochromatic, and their frequency is indeed
$\omega_{\rm o}$ (Fig.~\ref{fig3}(a)).
\par
\begin{figure}
\begin{center}
\leavevmode
\hbox{\epsfxsize=7cm \epsffile{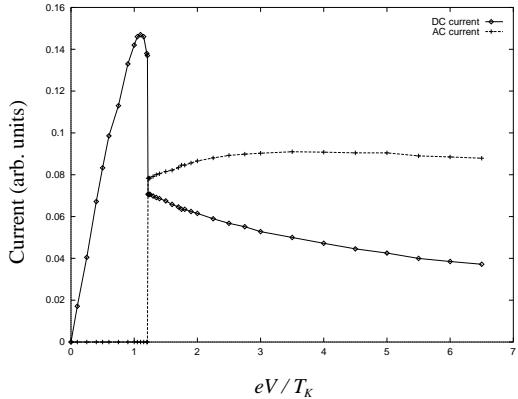}}
\end{center}
\caption{The direct and alternating parts of the current through the dot ($I_{\rm 
DC}$ and $I_{\rm AC}$)
as a function of voltage, $V$, in the case $\eta \approx 0.46$.}
\label{fig5}
\end{figure}

For the oscillating current $I(t)$,
one may identify $I_{\rm max}$ 
(the maximum instantaneous current) and $I_{\rm min}$ (the minimum);
one may thence define
the DC and AC parts: $I_{\rm DC} \equiv {1 \over 2} 
\left( I_{\rm min} +
I_{\rm max} \right)$; $I_{\rm AC} \equiv
{1 \over 2} \left( I_{\rm max} - I_{\rm min}
\right)$.  
These are plotted in Fig.~\ref{fig5} as functions 
of $V$.
The appearance of alternating and direct components to the current is a
manifestation of a two-frequency solution, for if 
the hybridization
fields have a two frequency 
form $\Gamma_{R} (t) =\Gamma_{L}^{*} (-t) = \Gamma_{+}e^{-i\omega_{\rm o}t} 
+\Gamma_{-}^{*}e^{i\omega_{\rm o}t}$,
then the current $I (t)= 
2J(1-\eta) \, \mbox{Im} \left(\Gamma_{R} (t)\Gamma_{L}^{*} (t)\right)$
takes the form
\begin{equation}
{I \over 2J(1-\eta)} = \mbox{Im} [\Gamma_{+}^{2}+\Gamma_{-}^{2}]+
2 \vert \Gamma_{+}\Gamma_{-}\vert 
\sin (2\omega_{\rm o}t + \phi ),
\end{equation}
where we have written $\Gamma_{+}\Gamma_{-}= \vert
\Gamma_{+}\Gamma_{-}\vert e^{-i\phi} $.
Phase coherence
between $\Gamma_{+}$ and $\Gamma_{-}$ is needed to 
stabilize the phase $\phi $ inside the oscillatory term.
\par
At small 
$V$, the $I{-}V$ curve is linear, 
and then peaks at $eV \sim k_B T_K$.
The behavior of the
differential conductance $G(V) \equiv dI_{\rm DC}/dV$ 
beyond the peak is novel, as it is affected by the sharp transition 
into the new regime where direct and alternating currents occur together.
As the voltage is increased from zero,
$G(V)$ changes sign at $eV \sim k_B T_K$, exhibits a
discontinuity at $V=V_c$, and eventually approaches zero 
from below.
\par
%
We conclude with a brief discussion on  the detection of 
this proposed oscillatory phenomenon.
In close analogy to the
Josephson effect, we expect that when the quantum dot is
exposed to radiation, the current-voltage profile will develop
Shapiro steps \cite{Shapiro:1963} when the oscillatory frequency is commensurate with the
frequency of the incident radiation. 
We give here a brief derivation of the expected effect. 
\par
A modulation in the bias voltage
$V (t)= V_{\rm o}+ V_{1}\sin \omega_{\rm in}t$
will lead to a modulation in the separation of the Kondo resonances
on opposite leads, given approximately
by $2\omega (t)=2\omega_{\rm o}+\frac{e V_{1}}{2\hbar }\cos
\omega_{\rm in}t$. 
Since this separation determines the
rate of change of the oscillatory current's phase, the current 
in the irradiated dot will
take the form
\begin{eqnarray}\label{}
I (t)&=& I_{\rm DC} + I_{\rm AC}\sin \left(2\int ^{t}\omega (t')dt' \right)\cr
&=& I_{\rm DC}+I_{\rm AC}\sin \left( 2\omega_{\rm o}t+ \frac{eV_{1}}{\hbar\omega_{\rm in}
}\sin (\omega_{\rm in}t)+\phi \right).
\end{eqnarray}
This function contains new Fourier components
at frequencies $2\omega_{\rm o}\pm n \omega_{\rm in}$, where $n$ is an integer,
so that when $2\omega_{\rm o}$ is a multiple of $\omega_{\rm in}$, additional
contributions appear in the direct current.
In a current biased quantum dot, 
we expect this rectification effect to lead to 
Shapiro steps in the current at voltages
where $\omega_{\rm in}$ is commensurate with the 
separation of the Kondo resonances $2\omega_{\rm o} (V)$. 
Experimentally, these steps should be visible in the range
$T_K<V_{\rm o}<U$, where $U$ is the charging energy of the dot.
Observation
of such Shapiro steps would constitute direct evidence of phase
coherent current oscillations in
the DC biased quantum dot.
\par
This research is partially supported by a US-Israel 
BSF grant and by the U.S.\ Department
of Energy.  One of us (CH)
is grateful to the Lindemann Trust Foundation and to
the EPSRC (UK) for financial
support.  The authors are pleased to acknowledge useful and
stimulating discussions with Drs
R.\ Aguado, L.\ Glazman, D.\ Langreth, O.\ Parcollet, and
A.\ Rosch.

\end{document}